\documentclass[conference]{IEEEtran}
\IEEEoverridecommandlockouts
\usepackage{cite}
\usepackage{amsmath,amssymb,amsfonts}
\usepackage{algorithmic}
\usepackage{graphicx}
\usepackage{textcomp}
\usepackage{xcolor}
\usepackage{listings}
\usepackage{color}
\def\BibTeX{{\rm B\kern-.05em{\sc i\kern-.025em b}\kern-.08em
    T\kern-.1667em\lower.7ex\hbox{E}\kern-.125emX}}

\begin{document}

\title{Applying Transparent Shaping for Zero Trust Architecture Implementation in AWS: A Case Study\\

}

\DeclareRobustCommand*{\IEEEauthorrefmark}[1]{%
  \raisebox{0pt}[0pt][0pt]{\textsuperscript{\footnotesize #1}}%
}
\author{\IEEEauthorblockN{Wenjia Wang\IEEEauthorrefmark{1},
Seyed Masoud Sadjadi\IEEEauthorrefmark{2},
Naphtali Rishe\IEEEauthorrefmark{3},
Arpan Mahara\IEEEauthorrefmark{4}
}
\IEEEauthorblockA{\textit{Knight Foundation School of Computing and Information Sciences} \\
\textit{Florida International University}\\
Miami, USA \\
Email:wwang048@fiu.edu\IEEEauthorrefmark{1},
sadjadi@cs.fiu.edu\IEEEauthorrefmark{2},
rishen@cs.fiu.edu\IEEEauthorrefmark{3},
amaha038@fiu.edu\IEEEauthorrefmark{4}
}

}

\maketitle

\begin{abstract}
\hspace{1cm} 
This study introduces a methodology integrating Zero Trust Architecture (ZTA) principles and Transparent Shaping into an AWS-hosted Online File Manager (OFM) application, enhancing security without substantial code modifications. We evaluate our approach with the Mozilla Observatory, highlighting significant security improvements and outlining a promising direction for applying Transparent Shaping and ZTA in cloud environments.

\end{abstract}

\section*{Nomenclature}
\addcontentsline{toc}{section}{Nomenclature}
Zero Trust Architecture (ZTA), Online File Manager (OFM), Amazon Web Service (AWS), Amazon Simple Storage Service (S3), Amazon Elastic Compute Cloud (EC2)
\section{Introduction}
Cloud computing has become integral to modern technology infrastructure in the digital age. The rapid adoption of cloud services, such as Amazon Web Services (AWS), has brought great convenience. While AWS offers numerous security features and best practices, the rapid pace of technological advancements and the ever-evolving threat landscape necessitates the continuous evaluation and improvement of security measures in cloud environments\cite{b1}. 

Traditionally, perimeter security has been the primary defense mechanism for organizations, employing firewalls and intrusion detection systems to protect their digital assets \cite{b2}. Now, Cybersecurity threats have become more advanced, targeted, and persistent, with attackers employing a wide range of techniques to infiltrate networks, steal sensitive data, and disrupt operations\cite{b3}\cite{b3_01}. Traditional perimeter security models are not sufficient to protect against these threats, particularly when they come from insiders or compromised accounts\cite{b4}. 

Instead, Zero Trust has emerged as a response to the limitations of traditional security models in addressing the evolving threat landscape and the increasing complexity of modern IT environments. It offers a more holistic, adaptive, and comprehensive approach to securing data, applications, and users, helping organizations better protect their assets and mitigate risks\cite{b4}\cite{b5}.

\subsection{Problem}

The growing adoption of AWS applications and environments has led to an increasing demand for robust security measures\cite{b6}. However, AWS's reliance on traditional perimeter security models poses a significant challenge in effectively implementing the Zero Trust model within its infrastructure; the perimeter-based approach is insufficient to address the evolving threat landscape, leaving AWS applications vulnerable to advanced attacks, insider threats, and other security risks\cite{b1}\cite{b6}. While organizations recognize the need to transition to a Zero Trust model, the lack of appropriate tools and methodologies has hindered progress in this direction. 

The Transparent Shaping model \cite{b7}, pioneered by Dr. Masoud Sadjadi and his colleagues, offers a novel approach for enhancing the adaptability, scalability, and security of existing software applications, particularly in distributed environments. By employing aspect-oriented techniques to separate crosscutting concerns from the main application logic, Transparent Shaping enables seamless integration of new features and behaviors without modifying the underlying source code\cite{b7}\cite{b8}.


This paper aims to explore the transition from perimeter security to Zero Trust Architecture (ZTA) in the context of existing AWS applications, emphasizing the importance of Transparent Shaping in enhancing cloud security. We delve into the key principles of Zero Trust, the benefits of applying this model to AWS applications, and the practical steps required for implementing transparent shaping. By doing so, we seek to provide a comprehensive understanding of the evolution and adoption of Zero Trust security in the realm of cloud computing, highlighting its significance in safeguarding the ever-expanding digital landscape.

\subsection{Contribution}

\begin{enumerate}
\item Enhanced Application Security: We introduce a methodology that merges ZTA principles with the Transparent Shaping model for AWS-hosted applications. This approach improves security without major code changes, addressing vulnerabilities like weak passwords and risks from file uploads, and is easily replicable.
\item Custom ZTA Strategy: Our customized ZTA strategy for OFM showcases the practical implementation of Zero Trust in AWS, highlighting policy formulation and risk mitigation.

\item Research Foundation: Our findings provide a basis for further study on Transparent Shaping and Zero Trust in cloud environments, promising advancements in cloud application security.

\item Research Experiment Series: This paper initiates a sequence of studies on applying Transparent Shaping and ZTA in AWS, aiming to enrich cloud security knowledge and practices.

\end{enumerate}

\section{Background and Related Work}

This section provides an overview of the transition from perimeter security to the Zero Trust model, emphasizing their application and challenges in cloud environments. Our aim is to offer a detailed understanding of these two approaches, highlighting their strengths and weaknesses. We start by discussing the traditional method of Perimeter Security and its evolution into the increasingly prevalent Zero Trust strategy. The shift towards more robust and flexible security measures is considered in light of recent technological developments and the changing cyber threat landscape. This discussion extends into the exploration of the cybersecurity aspects of cloud environments, referencing significant studies that address diverse challenges in this evolving field.

\subsection{From Perimeter Security to Zero Trust}

Perimeter Security, otherwise known as, the ``castle-and-moat", focuses on creating a secure boundary around an organization's network, typically using firewalls, intrusion detection systems, and intrusion prevention systems\cite{b9}. This approach assumes that once users and devices are authenticated and granted access to the network, they are trustworthy. However, perimeter security can be vulnerable to insider threats, sophisticated attacks that bypass security measures, and the increasing reliance on cloud services, which blurs the traditional network boundary\cite{b10}.

In contrast, Zero Trust operates under the principle of ``never trust, always verify"\cite{b11}. It assumes that no user or device is inherently trustworthy, regardless of whether they are inside or outside the network perimeter. Zero Trust focuses on granular access control, strong authentication, microsegmentation, continuous monitoring, and the least privilege principle. This approach helps protect sensitive data by minimizing the attack surface and reducing the potential for lateral movement within networks\cite{b12}. Additionally, compliance is complex yet increasingly important \cite{b12_01}, and adopting Zero Trust is a strategic approach to ensure applications are compliant and adaptive to evolving digital threats \cite{b12_02}.

The shift from perimeter security to Zero Trust has been driven by the need for a more comprehensive and adaptive security model that addresses the limitations of traditional perimeter-based approaches. Over the past decade, a growing body of research has emerged, focusing on the implementation and application of Zero Trust principles to various aspects of information technology\cite{b13}.

Palo Alto Networks was among the first to adopt the Zero Trust concept introduced by John Kindervag in 2010\cite{b4}. Since then, numerous studies have been conducted to explore the practical applications of Zero Trust. National Institute of Standards and Technology (NIST) has also published a comprehensive guide on ZTA, providing a framework for organizations to design and deploy Zero Trust solutions\cite{b5}.
Following this whitepaper, NIST released the ``NIST SPECIAL PUBLICATION 1800-35E: Implementing a Zero
Trust Architecture" in December 2020. This publication offers an in-depth exploration of implementing ZTA, delving into the practical application of Zero Trust principles. It presents various use cases, best practices, and detailed guidance for organizations transitioning from traditional perimeter-based security models to a more robust and adaptive Zero Trust approach
\cite{b5_1}.

In recent years, researchers have started exploring the application of Zero Trust principles to specific areas of technology, such as cloud computing and the Internet of Things (IoT). T. Dimitrakos et al. applied Zero Trust model to consumer IoT\cite{b14}. S. Mehraj and M. T. Banday successfully established Zero Trust strategy in the cloud computing environment\cite{b15}. M. K. Pratt investigated the challenges and best practices associated with implementing Zero Trust for remote access, providing insights into how organizations can transition from traditional VPN-based solutions to more secure Zero Trust-based remote access\cite{b16}. S. Teerakanok, T. Uehara, and A. Inomata have reviewed challenges and general steps to consider when migrating from legacy architecture to ZTA\cite{b17}.

These studies collectively demonstrate the growing importance of Zero Trust in the information security landscape and the increasing need for effective tools and methodologies to facilitate its implementation across different technologies and environments. The existing literature on Zero Trust security provides valuable insights into the development and application of this paradigm. Incorporating Zero Trust principles into the Transparent Shaping model and applying them to AWS applications can help further enhance security in cloud environments.

\subsection{Cybersecurity Landscape of Cloud Environments}
As we delve deeper into the digital age, the cybersecurity landscape of cloud environments becomes increasingly complex, necessitating vigilant examination and constant adaptation. This section serves to examine the intricacies of this evolving field, encapsulating a range of studies that shed light on a multitude of issues.

Fischer et al. \cite{b171} presented data flow authentication (DFAuth) as a solution for encrypted computation, ensuring data security in cloud environments. To detect and prevent intrusions, a distributed taint tracking system was proposed\cite{b172} to monitor information flows among multiple hosts in the cloud. Besides technical security solutions, user awareness and understanding of cloud services were studied\cite{b173}, emphasizing the need for improved user education regarding cloud security. Furthermore, misconfigurations and codebase vulnerabilities are significant threats in cloud environments. To address this, one study presented a novel framework\cite{b174} to secure configuration files against vulnerabilities and compliance issues, while another introduced SWAN\cite{b175}, a machine-learning approach to detect sources, sinks, validators, and sanitizers in Java programs. Lastly, a method for the automated monitoring of code repositories was proposed\cite{b176} to identify and implement security patches swiftly.

Public clouds, despite the advancements in hypervisors and containerization frameworks, are still vulnerable to co-residency attacks, as shown in \cite{b177}. The study highlighted the necessity of addressing co-residency attacks on containers running on modern orchestration systems. Furthermore, with the rise of electronic consent (e-consent), privacy and security challenges have arisen. The review in \cite{b178} outlined these challenges, emphasizing the need for security- and privacy-by-design techniques in e-consent platforms.

Implementations of network protocols are often prone to vulnerabilities due to developer mistakes. \cite{b179} proposed a combination of fuzzing and symbolic execution to find security vulnerabilities in network protocol implementations, which could be beneficial in preventing attacks through network protocols.

Meanwhile, ensuring the security of enclaved execution on small microprocessors poses another challenge\cite{b1710}. The paper proposed a design for interruptible enclaves that satisfies a general formal criterion for the security of a processor extension, based on the concept of full abstraction from programming languages.

Microarchitectural vulnerabilities are another significant concern, as the Spectre attacks have shown. Paper\cite{b1711} introduced a comprehensive formal microarchitectural model capable of representing out-of-order and speculative behavior in high-performance pipelined architectures. It also discovered potential vulnerabilities and analyzed the effectiveness of proposed countermeasures.

Additionally, microarchitectural timing covert channels can be exploited in Infrastructure as a Service (IaaS) clouds, as demonstrated in paper\cite{b1712}. Using processor memory order buffer (MOB), a robust covert communication channel was built, which could be used for multi-tenant detection in IaaS clouds. To increase trust in cloud platforms, \cite{b1713} proposed an approach for configuration validation. This method provides transparency regarding the actual fulfillment of service requirements, enhancing user trust in cloud services.

\subsection{AWS Security Practices and Challenges}
AWS has become the largest cloud service provider, offering a wide array of services to businesses and individuals worldwide\cite{b18}. Given the increasing reliance on cloud computing and the growing complexity of cloud-based architectures, several studies have explored AWS and its security mechanisms to ensure data protection, privacy, and compliance.

In this section, we review the literature on AWS security, covering essential security features, practices, and the challenges associated with securing AWS applications.

\subsubsection{AWS Shared Responsibility Model}
AWS operates under a shared responsibility model, where the platform is responsible for the security of the cloud, while customers are responsible for security in the cloud\cite{b19}. This model delineates the respective security roles and responsibilities of AWS and its customers, ensuring that both parties contribute to maintaining a secure cloud environment.

\subsubsection{AWS Security Services}AWS provides various security services to help customers secure their applications and data, such as Identity and Access Management (IAM), Amazon GuardDuty, Amazon Inspector, AWS Security Hub, and AWS Shield \cite{b20}. These services cover a wide range of security concerns, including access control, intrusion detection, vulnerability assessment, compliance monitoring, and distributed denial-of-service (DDoS) protection.

\subsubsection{AWS Security Compliance}Compliance is a critical aspect of cloud security, and AWS addresses this concern by offering various compliance certifications and adherence to regulatory standards \cite{b21}\cite{b22}. However, customers must ensure that their applications and data remain compliant within their specific industry and regional regulations.

\subsubsection{AWS Security Practices}Several studies and whitepapers have discussed best practices for securing AWS applications, focusing on topics like least privilege, encryption, logging, monitoring, and incident response\cite{b23}\cite{b24}. These practices are designed to help customers effectively manage their security responsibilities and minimize potential risks.

\subsubsection{Security Challenges in AWS}
As organizations increasingly adopt AWS applications and environments, understanding the security challenges associated with this technology has become a critical area of research. Various studies have focused on identifying the key security concerns and proposing solutions to mitigate risks in AWS applications and environments.

One of the earliest comprehensive studies on AWS security was conducted by Modi et al., who provided an in-depth analysis of the security issues faced by cloud consumers and service providers\cite{b25}. They emphasized the need for securing data in transit and at rest, as well as the importance of access control and identity management.

All these related work in AWS security challenges underscores the importance of adopting comprehensive security measures to protect cloud resources and services. By understanding the specific risks and vulnerabilities associated with AWS, organizations can better secure their applications and environments against a wide range of cyber threats.

\subsection{Security Enhancements in Cloud Networks using Zero Trust}

The growing need for more secure cloud environments has led researchers to investigate the application of Zero Trust principles in cloud networks. In this section, we review recent advancements in Zero Trust security applied to cloud networks.

DeCusatis et al. presented a zero trust network security architecture that leverages first-packet-based authentication and a steganographic overlay to provide enhanced security for SDN controllers against cyberattacks in both enterprise and cloud computing environments\cite{b25_2}. Eidle et al. presented an autonomic security system aligned with Zero Trust principles, proactively defending against threats from both inside and outside the network, and the system has the potentiality to be applied to cloud-based platforms\cite{b25_3}. Mandal et al. proposed a zero trust access control policy to protect enterprise cloud resources\cite{b25_4}. \cite{b25_5} evaluated the performance impact of implementing Zero Trust security using the Istio service mesh in multi-cloud deployments, finding minimal overhead in terms of latency and resource usage.

 A Zero Trust Federated Identity and Access Management framework for cloud-based computing environments has been propose in \cite{b25_51}, and it can be used to prevent unauthorized access to customer digital assets placed under a CSP’s management. \cite{b25_511} presented an online model checking method for zero trust security policies that ensures consistency in policy implementation and dynamically detects policy compliance with system security protocols through pre-model and post-model detection techniques. \cite{b25_52} discusses the importance of implementing a strong security posture in the cloud and how to effectively combine traditional security measures with ZTA. Additionally, \cite{b25_6} provides a comparative review of recently published zero-trust-based cloud network models, frameworks, and proof-of-concept employed for network security to help organizations understand the latest developments in this field.

The adoption of Zero Trust in cloud networks, as demonstrated in the case studies presented in this section, has proven to be a valuable and indispensable strategy for organizations aiming to safeguard their digital assets and ensure the ongoing security of their systems.

\section{Toward Transparent Shaping}
Introduced by Dr. Masoud Sadjadi \cite{b7}, the Transparent Shaping model efficiently negotiates non-functional factors such as security, performance, and adaptability in current software applications, particularly in the realms of distributed and cloud computing. Here, we elucidate the salient features that make Transparent Shaping an excellent choice for adopting the Zero Trust model within AWS applications. In this section, we discuss why it is a potent technique that can adeptly navigate the intricacies of infusing Zero Trust principles into pre-existing AWS applications, while simultaneously addressing the heterogeneous security facets of AWS environments.

\begin{itemize}
\item Separation of concerns: One of the significant advantages of Transparent Shaping lies in its distinctive separation of concerns. It distinguishes between functional aspects (such as application logic) and non-functional facets (such as security), utilizing aspect-oriented techniques. This demarcation allows for a smooth incorporation of Zero Trust principles like continuous authentication and least privilege access, without necessitating modifications to the core source code of AWS applications. Consequently, it ensures minimum disruption to established systems.

\item Prompt integration of security mechanisms: Transparent Shaping enables the swift integration of advanced security mechanisms and performance optimizations into AWS applications. By adopting Transparent Shaping, developers can easily implement Zero Trust policies, such as micro-segmentation and real-time monitoring, which are crucial for protecting against modern threats in a cloud environment.

\item Scalability and adaptability: Transparent Shaping champions scalability and adaptability, making it an ideal approach for AWS applications that typically operate in a dynamic and ever-changing environment. With the Zero Trust model's emphasis on continuous monitoring and evaluation, Transparent Shaping ensures that AWS applications can quickly adapt to evolving security requirements and respond to potential threats effectively.

\item Reduced development and maintenance efforts: By employing Transparent Shaping, developers can integrate Zero Trust principles into AWS applications with reduced development and maintenance efforts. This strategy optimizes the process of upgrading application security, enabling organizations to adopt Zero Trust more efficiently and in a more cost-effective manner.
\end{itemize}

Hence, Transparent Shaping offers a robust and flexible approach for implementing the Zero Trust model in AWS applications. Its ability to separate functional and non-functional concerns, along with its emphasis on rapid integration, scalability, and adaptability, makes it an ideal method for enhancing the security of AWS applications in the context of the Zero Trust model.

\section{Experiment}
In this section, we present the step-by-step process of incorporating Transparent Shaping and Zero Trust tenets in an existing AWS application: Online File Manager (OFM). OFM is an open-source web storage solution allowing authenticated users to store, access, delete, and share files\cite{b26}. The application is built using React for the frontend and integrates with AWS services for backend functionality.

\subsubsection{Implementing} In the first step of our experiment, we deployed the OFM project in AWS by utilizing the provided source code. The project deployed is accessible via the following URL: https://main.d2e9j6uk04urb2.amplifyapp.com/. This initial implementation serves as the basis for our subsequent experiments. The OFM is designed using AWS Amplify, a set of tools and services that enable front-end web and mobile developers to build full-stack applications on AWS more efficiently\cite{b40}.

\subsubsection{Analyzing}
Built on AWS, the OFM platform employs Amazon Cognito for user authentication, Amazon Simple Storage Service (S3) for file storage, Amazon Route 53 for domain configuration, and Amazon API Gateway for API management. We visualize all the AWS services that OFM currently uses in Fig. 1.
\begin{figure}
\centering
\includegraphics[trim={5cm 6cm 3cm 3.2cm},clip, width=15cm,keepaspectratio]{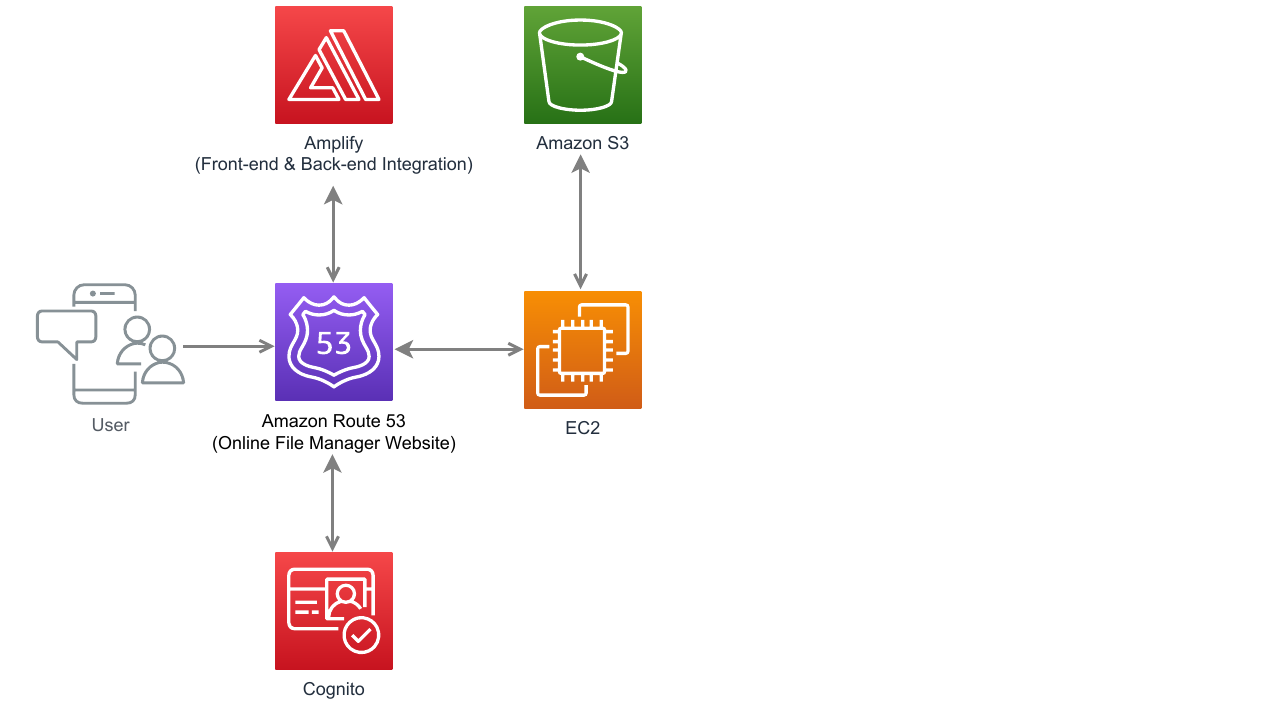}
\caption{Original Architecture of ``Online File Manager"}
\end{figure}

\begin{figure}
    \centering
    \includegraphics[width=88mm,]{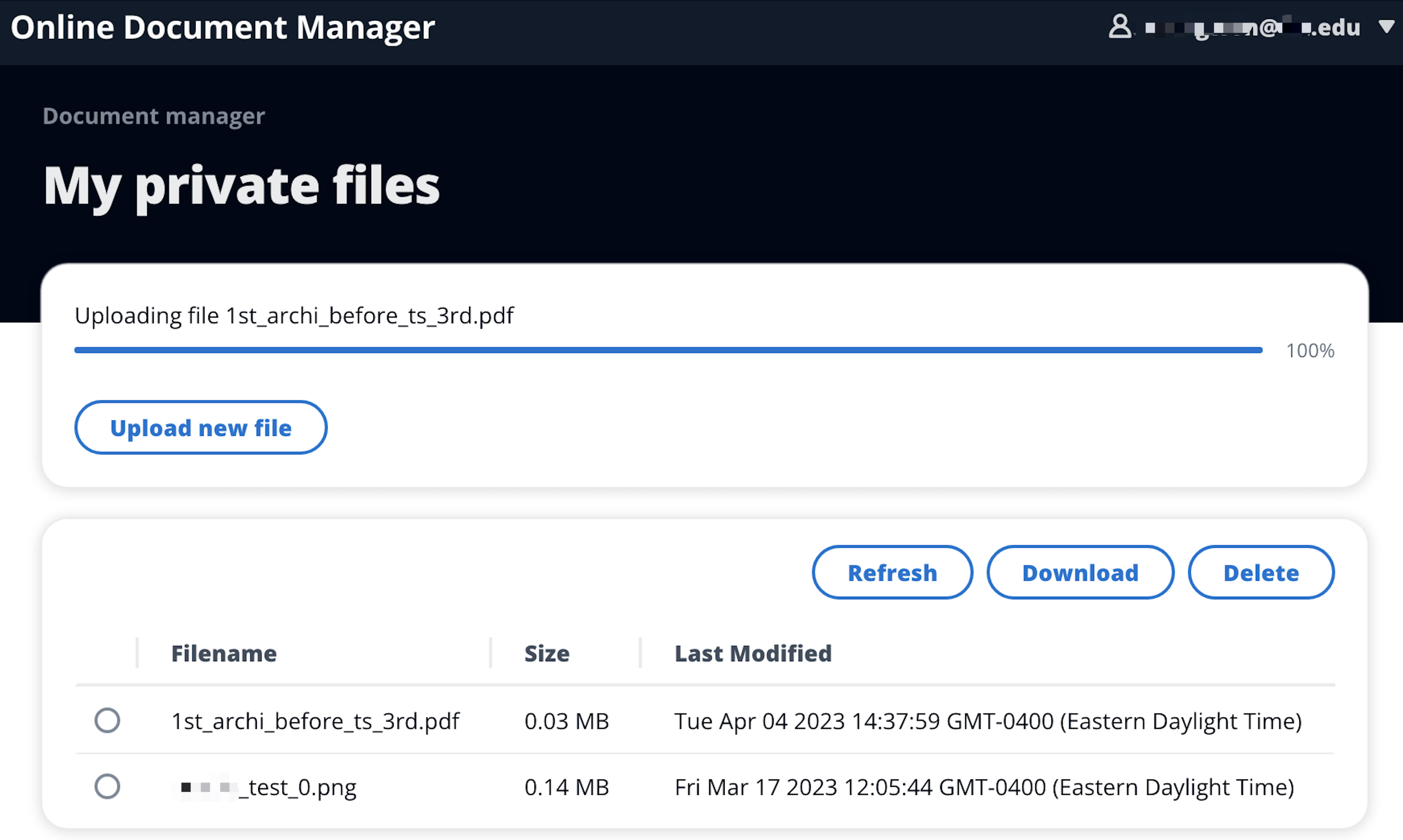}
    \caption{Online File Manager (OFM)}
    \label{fig:my_label}
\end{figure}

\subsubsection{Scanning}
We then use the Mozilla Observatory web security scanner to assess the current security posture of the application and identify vulnerabilities. The initial scanning results are presented in Table I.
Mozilla Observatory is an open-source tool, developed by Mozilla that evaluates a range of security features and settings, producing a comprehensive report with a letter-grade score and recommendations for improvement\cite{b27}.
\begin{table*}[t]\centering

    \centering
    \caption{SCANNING RESULTS BEFORE IMPLEMENTING Zero Trust ARCHITECTURE}
    \begin{tabular}{|l|l|l|l|}
    \hline
        Test & Pass & Score & Reason \\ \hline
        Content Security Policy & $\times$ & -25 & CSP header not implemented \\ \hline
        Cookies & -- & 0 & No cookies detected \\ \hline
        Cross-origin Resource Sharing & 
        \checkmark & 0 & Content is not visible via cross-origin resource sharing (CORS) files or headers \\ \hline
        HTTP Public Key Pinning & -- & 0 & HTTP Public Key Pinning (HPKP) header not implemented (optional) \\ \hline
        HTTP Strict Transport Security & 
        $\times$ & -20 & HTTP Strict Transport Security (HSTS) header not implemented \\ \hline
         Redirection & \checkmark & 0 & Initial redirection is to HTTPS on same host, final destination is HTTPS \\ \hline
        Referrer Policy & -- & 0 & Referrer-Policy header not implemented (optional) \\ \hline
        Subresource Integrity & -- & 0 & Subresource Integrity (SRI) not implemented, but all scripts are loaded from a similar origin \\ \hline
        X-Content-Type-Options & 
        $\times$ & -5 & Not implemented \\ \hline
        X-Frame-Options & $\times$ & -20 & Not implemented \\ \hline
        X-XSS-Protection & $\times$ & -10 & Not implemented \\ \hline
    \end{tabular}
\end{table*}

\subsubsection{Separating Functional and Non-functional Concerns}

Within the Transparent Shaping framework, we acknowledge the distinction between functional concerns (such as user authentication, file storage, and sharing) and non-functional concerns (including security, performance, scalability, and availability). This distinction is crucial for applying Zero Trust principles effectively, ensuring a robust security posture without compromising the application's functionality or user experience. While our implementation focuses on practical security enhancements, we recognize the theoretical importance of separating these concerns in line with Transparent Shaping principles, aiming for an application that is secure, performant, and scalable.
\subsubsection{Incorporating Zero Trust Principles}
We design a ZTA implementation plan for the non-functional concerns of OFM based on the standard implementation procedure made by NIST\cite{b5}.

Step 1: Identify Actors on OFM. There are three types of Actors on OFM:

\begin{itemize}
    \item End-users: Individuals use the OFM application to store, access, and share files. They interact with the system through authentication and authorization processes, file storage and retrieval, file-sharing features, etc.
    \item
Administrators or Developers: These actors are responsible for designing, implementing, maintaining, and managing the OFM application. They have access to the source code, development environments, and various AWS services used in the application.
    \item 
AWS services: Various AWS services employed in the OFM application, including AWS Route 53, AWS S3, AWS EC2, etc. They interact with each other and the application to perform their designated functions.
\end{itemize}

Step 2: Identify Assets Owned by the OFM. Critical assets include user accounts, Amazon S3 buckets, Amazon Route 53, application source code, and infrastructure configuration.

Step 3: Identify Key Processes and Evaluate Risks Associated with Executing Processes. We examined processes such as user authentication, file management, and sharing. Vulnerabilities identified included not just insufficient password policies, lack of file type control, and susceptibility to malware or ransomware, but also specific security risks highlighted by Mozilla Observatory scanning. These encompassed the absence of Content Security Policy (CSP), inadequate HTTP Strict Transport Security (HSTS), and missing X-Content-Type-Options, X-Frame-Options, and X-XSS-Protection headers, presenting significant risks to web application security.

Step 4: Formulating Policies and Solutions for OFM potential risks. To mitigate these risks, we:
\begin{itemize}
\item Enhanced password policies for greater security.
\item Implemented file format and size checks for upload validation.
\item Introduced anti-malware scanning for uploaded files.
\item Addressed Mozilla Observatory's findings by updating AWS Amplify build settings to include critical security headers: enforcing CSP for script control, enabling X-Content-Type-Options to prevent MIME type sniffing, setting X-Frame-Options to guard against clickjacking, and activating X-XSS-Protection for additional XSS attack prevention.
\end{itemize}

Step 5: Initial deployment and testing. Deploy the chosen ZTA components and implement the developed policies to OFM. After the deployment, conduct testing to validate the effectiveness of the implemented solutions and to identify any unforeseen issues that might arise due to the changes. 

\begin{table*}[t]\centering

    \centering
    \caption{SCANNING RESULTS AFTER IMPLEMENTING Zero Trust ARCHITECTURE}
    \begin{tabular}{|l|l|l|l|}
    \hline
        Test & Pass & Score & Reason \\ \hline
        Content Security Policy & \checkmark & +5 & Content Security Policy (CSP) implemented without 'unsafe-inline' or 'unsafe-eval' \\ \hline
        Cookies & -- & 0 & No cookies detected \\ \hline
        Cross-origin Resource Sharing & 
        \checkmark & 0 & Content is not visible via cross-origin resource sharing (CORS) files or headers \\ \hline
        HTTP Public Key Pinning & -- & 0 & HTTP Public Key Pinning (HPKP) header not implemented (optional) \\ \hline
        HTTP Strict Transport Security & 
        $\times$ & -20 & HTTP Strict Transport Security (HSTS) header not implemented \\ \hline
         Redirection & \checkmark & 0 & Initial redirection is to HTTPS on same host, final destination is HTTPS \\ \hline
        Referrer Policy & -- & 0 & Referrer-Policy header not implemented (optional) \\ \hline
        Subresource Integrity & -- & 0 & Subresource Integrity (SRI) not implemented, but all scripts are loaded from a similar origin \\ \hline
        X-Content-Type-Options & 
        \checkmark & 0 & X-Content-Type-Options header set to "nosniff" \\ \hline
        X-Frame-Options & \checkmark & 0 & X-Frame-Options (XFO) header set to SAMEORIGIN or DENY \\ \hline
        X-XSS-Protection & \checkmark & 0 & Deprecated X-XSS-Protection header set to "1; mode=block"	
 \\ \hline
    \end{tabular}
\end{table*}
\subsubsection{Adapting OFM Application Utilizing Transparent Shaping Model}

Following ZTA implementation strategy outlined in Step 5, we applied comprehensive security measures across the OFM application's lifecycle. This entailed critical updates to AWS Amplify's build settings, exemplifying the Transparent Shaping model's integration at the infrastructure level. Our modifications aimed to establish a robust security policy throughout the app's deployment phase.

Key enhancements were made to the build specification YML file of AWS Amplify, incorporating custom headers to bolster defenses against prevalent web vulnerabilities, directly reflecting our formulated policies and solutions. These headers are:
\begin{itemize}
    \item Content-Security-Policy: Enforces a strict policy that prevents the execution of unauthorized scripts, effectively guarding against Cross-Site Scripting (XSS) attacks.
    \item X-Content-Type-Options: Prevents the browser from interpreting files as a different MIME type than what is specified by the content type in the HTTP headers, mitigating MIME type security risks.
    \item X-Frame-Options: Protects against clickjacking attacks by prohibiting the application from being embedded within an iframe.
    \item X-XSS-Protection: Activates browser mechanisms to reconfigure the page in response to detected XSS attacks.
\end{itemize}

These headers, adhering to the "never trust, always verify" principle of Zero Trust, ensure thorough validation of all resources, scripts, and frames, significantly reducing the application's vulnerability surface.

To further align with the policies and solutions identified in Step 5, we introduced the ``TransparentShapingWrapper" around the ``UploadFileCard" component. This adaptation not only addresses the need for secure file handling, as indicated by our risk assessment, but also seamlessly integrates additional security measures without altering the core application code. This wrapper enforces pre-processing checks for file size and format, along with post-processing PDF sanitization, ensuring all uploads comply with our security standards. The adoption of a strengthened password policy further secures user accounts against unauthorized access.

For a comprehensive view of the code structure, the changes implemented, and the entire progression of the Transparent Shaping model integration, please visit our GitHub repository\cite{b51}, where we provide detailed documentation and source code.

\subsubsection{Evaluating the Security Improvements}

To evaluate the impact of these infrastructure-level security enhancements, alongside the application-level improvements introduced by Transparent Shaping and ZTA, we conducted a  security assessment using Mozilla Observatory again. The aim was to measure the security posture of the OFM application before and after these modifications, the scanning result after are presented in Table II.

\section{Results}
The implementation of ZTA principles through the Transparent Shaping model within OFM has demonstrated a significant advancement in securing cloud-based applications hosted on AWS. The application of Transparent Shaping as a pathway to achieving ZTA has yielded measurable improvements in the application's security posture, as evidenced by the updated scanning results using Mozilla Observatory shown in Tables I \& II.
    
Enhancements Observed:
\begin{itemize}

   \item  Implementing controls on file type and size directly mitigates risks from malicious uploads, showcasing Transparent Shaping's ability to enforce crucial input validations seamlessly.
    \item Content Security Policy (CSP) Implementation: The activation of CSP as part of the AWS Amplify build settings adjustment represents a critical step forward in protecting the application against cross-site scripting (XSS) and other injection attacks. This policy restricts the sources from which scripts can be loaded, effectively reducing potential attack vectors.
    
    \item Robust HTTP Headers Integration: The inclusion of HTTP headers such as X-Content-Type-Options and X-Frame-Options enhances the application's resilience against MIME type sniffing and clickjacking attacks, respectively. These measures exemplify the application of ZTA principles at the infrastructure level, ensuring a base level of security across all application interactions.
 
\end{itemize}

Transparent Shaping enabled smooth integration of ZTA principles into OFM, demonstrating how to bolster cloud application security without extensive redevelopment. This highlights Transparent Shaping's role in implementing ZTA efficiently, preserving application performance and user experience while improving security.

The results of this case study validate the effectiveness of combining Transparent Shaping with ZTA to secure cloud-based applications. By documenting the integration process and the subsequent improvements in security posture, this research contributes valuable insights into the application of ZTA in cloud environments, particularly for applications hosted on AWS.

\section{Conclusion}
In this paper, Our study effectively integrates ZTA with the Transparent Shaping model in an AWS-hosted application, markedly improving its security. Strategic updates to AWS Amplify build settings and the adoption of advanced security measures addressed key vulnerabilities, verified by Mozilla Observatory scans. This approach not only fortified the application against threats but also preserved its functionality and user experience. Implementing ZTA through Transparent Shaping proves to be a viable method for enhancing cloud application security, offering insights for future research and potential broad application in cybersecurity practices.

\section{Future Work}

The experiment yielded encouraging outcomes, but further research is required. Specifically, future work should focus on the development or integration of comprehensive security analysis tools for ZTA, as this was a significant limitation in the present study.

This investigation is just the first in a series examining the integration of Transparent Shaping and Zero Trust in AWS environments. Future studies will refine the tools and methods employed and will deepen our understanding of these principles' practical implications.

An important goal for future research is to automate the process of implementing Zero Trust and Transparent Shaping. The current study involved a manual implementation, but for scalability, automatic application to new or updated code is desirable. This will enhance the efficiency and accessibility of these security measures in AWS applications.

\section*{Acknowledgment}
This material is based in part upon work supported by the National Science Foundation under Grant No.  MRI20 CNS-2018611.


\end{document}